\documentstyle[ltwol,epsfig,psfig]{article}
\newcommand {\ignore}[1]{}

\def\21{$SU(2) \ot U(1)$}
\def\321{$SU(3) \ot SU(2) \ot U(1)$}

\def\pr#1#2#3{           { Phys. Rev. }{\bf #1}, #3 (19#2)}

\def\prl#1#2#3{          { Phys. Rev. Lett. }{\bf #1}, #3 (19#2)}

\def\n.c.#1#2#3{         { Nuovo Cim. }{\bf #1}, #3 (19#2)}
\def\r.n.c.#1#2#3{       { Riv. del Nuovo Cim. }{\bf #1}, #3 (19#2)}

\begin{document}
\title{ANOMALOUS HIGGS COUPLINGS AT COLLIDERS}
\author{M.\ C.\ Gonzalez-Garcia}
\address{Instituto de F\'{\i}sica Corpuscular - C.S.I.C./Univ. de Val\`encia.
46100 Burjassot, Val\`encia, SPAIN }
\twocolumn[\maketitle
\abstracts{I summarize our results on 
the attainable limits on the coefficients of
dimension--6 operators from the analysis of Higgs boson
phenomenology using data taken at Tevatron RUNI and LEPII. 
Our results show that the coefficients of Higgs--vector boson
couplings can be determined with unprecedented accuracy. 
Assuming
that the coefficients of all ``blind'' operators are of the same
magnitude, we are also able to impose bounds on
the anomalous vector--boson triple couplings comparable to those 
from double gauge boson production at the Tevatron and LEPII}.
] 

\section{Introduction}
\label{intr}
Despite the impressive agreement of the Standard Model (SM)
predictions for the fermion--vector boson couplings with the
experimental results, the couplings among the gauge bosons are
not determined with the same accuracy. The $SU_L(2) \times
U_Y(1)$ gauge structure of the model completely determines these
self--couplings, and any deviation can indicate the existence of
new physics beyond the SM. 

Effective Lagrangians are useful to describe and explore the
consequences of new physics in the bosonic sector of the SM
\cite{classical,linear,drghm,hisz}. After integrating out the
heavy degrees of freedom, anomalous effective operators can
represent the residual interactions between the light states. 
Searches for deviations on the
couplings $WWV$ ($V = \gamma, Z$) have been carried out at
different colliders and recent results \cite{fermilab} include
the ones by CDF \cite{CDF}, and D\O ~Collaborations \cite{D0,D02}.
Forthcoming perspectives on this search at LEP II CERN Collider
\cite{lep2,snow}, and at upgraded Tevatron Collider
\cite{tevatron} were also reported. 

In the framework of effective Lagrangians respecting the local
$SU_L(2) \times U_Y(1)$ symmetry linearly realized,  the
modifications of the couplings of the Higgs field ($H$) to the
vector gauge bosons ($V$) are related to the anomalous triple
vector boson vertex \cite{linear,drghm,hisz,hsz}. 
Here, I summarize our results on the attainable limits on the 
coefficients of
dimension--6 operators from the analysis of Higgs boson
phenomenology using data taken at Tevatron RUNI and LEPII. 
Our results show that the coefficients of Higgs--vector boson
couplings can be determined with unprecedented accuracy. 
Assuming
that the coefficients of all ``blind'' operators are of the same
magnitude, we are also able to impose bounds on
the anomalous vector--boson triple couplings comparable to those 
from double gauge boson production at the Tevatron and LEPII.

\section{Effective Lagrangians}
\label{eff:hig}
A general set of dimension--6 operators that involve  gauge
bosons and the Higgs scalar field, respecting local $SU_L(2)
\times U_Y(1)$ symmetry, and $C$ and $P$ conserving, contains
eleven  operators \cite{linear,drghm}. Some of these operators
either affect only the Higgs self--interactions or contribute to
the gauge boson two--point functions at tree level and can be
strongly constrained from low energy physics below the present
sensitivity of high energy experiments \cite{drghm,hisz}. The
remaining five ``blind'' operators can be written as
\cite{linear,drghm,hisz},
\begin{eqnarray}
&& {\cal L}_{\mbox{eff}} = \sum_i \frac{f_i}{\Lambda^2} {\cal O}_i 
= \frac{1}{\Lambda^2} \Bigl[ 
f_{WWW}\, Tr[\hat{W}_{\mu \nu}\hat{W}^{\nu\rho}\hat{W}_{\rho}^{\mu}] 
\nonumber \\
&&+ f_W (D_{\mu} \Phi)^{\dagger} \hat{W}^{\mu \nu} (D_{\nu} \Phi) 
+ f_B (D_{\mu} \Phi)^{\dagger} \hat{B}^{\mu \nu} (D_{\nu} \Phi) 
\\ 
&&+ f_{WW} \Phi^{\dagger} \hat{W}_{\mu \nu} \hat{W}^{\mu \nu} \Phi  
+ f_{BB} \Phi^{\dagger} \hat{B}_{\mu \nu} \hat{B}^{\mu \nu} \Phi 
  \Bigr] 
\nonumber
\end{eqnarray}
where $\Phi$ is the Higgs field doublet, and 
\begin{displaymath}
\hat{B}_{\mu\nu} = i (g'/2) B_{\mu \nu} \;\;\hat{W}_{\mu \nu} = i (g/2)
\sigma^a W^a_{\mu \nu}
\end{displaymath}
with $B_{\mu \nu}$ and $ W^a_{\mu \nu}$ being the field strength
tensors of the $U(1)$ and $SU(2)$ gauge fields respectively. 

Anomalous $H\gamma\gamma$, $HZ\gamma$, and $HZZ$ and $HWW$ and 
couplings are generated by (1), which, in the unitary gauge, are
given by
\begin{eqnarray}
{\cal L}_{\mbox{eff}}^{\mbox{H}} &=& 
g_{H \gamma \gamma} H A_{\mu \nu} A^{\mu \nu} + 
g^{(1)}_{H Z \gamma} A_{\mu \nu} Z^{\mu} \partial^{\nu} H \nonumber \\ 
&+& g^{(2)}_{H Z \gamma} H A_{\mu \nu} Z^{\mu \nu}
+ g^{(1)}_{H Z Z} Z_{\mu \nu} Z^{\mu} \partial^{\nu} H \nonumber \\
&+& g^{(2)}_{H Z Z} H Z_{\mu \nu} Z^{\mu \nu} +
g^{(2)}_{H W W} H W^+_{\mu \nu} W_{-}^{\mu \nu} \; \nonumber \\
&+&g^{(1)}_{H W W} \left (W^+_{\mu \nu} W_{-}^{\mu} \partial^{\nu} H 
+h.c.\right)\,
\label{H} 
\end{eqnarray}
where $A(Z)_{\mu \nu} = \partial_\mu A(Z)_\nu - \partial_\nu
A(Z)_\mu$. The effective couplings $g_{H \gamma \gamma}$,
$g^{(1,2)}_{H Z \gamma}$, and $g^{(1,2)}_{H Z Z}$  and 
$g^{(1,2)}_{H WW}$ are related
to the coefficients of the operators appearing in (1)
through,
\begin{eqnarray}
g_{H \gamma \gamma} &=& - \left( \frac{g M_W}{\Lambda^2} \right)
                       \frac{s^2 (f_{BB} + f_{WW})}{2} \; , 
\nonumber \\
g^{(1)}_{H Z \gamma} &=& \left( \frac{g M_W}{\Lambda^2} \right) 
                     \frac{s (f_W - f_B) }{2 c} \; ,  
\nonumber \\
g^{(2)}_{H Z \gamma} &=& \left( \frac{g M_W}{\Lambda^2} \right) 
                      \frac{s [2 s^2 f_{BB} - 2 c^2 f_{WW}]}{2 c}  \; , 
\label{g} \\ 
g^{(1)}_{H Z Z} &=& \left( \frac{g M_W}{\Lambda^2} \right) 
	              \frac{c^2 f_W + s^2 f_B}{2 c^2} \nonumber \; , \\
g^{(2)}_{H Z Z} &=& - \left( \frac{g M_W}{\Lambda^2} \right) 
  \frac{s^4 f_{BB} +c^4 f_{WW} }{2 c^2} \nonumber \; ,\\  
g^{(1)}_{H W W} &=& \left( \frac{g M_W}{\Lambda^2} \right) 
	              \frac{f_{W}}{2} \nonumber \; , \\
g^{(2)}_{H W W} &=& - \left( \frac{g M_W}{\Lambda^2} \right) 
  f_{WW} \nonumber \; ,  
\end{eqnarray}
with $g$ being the electroweak coupling constant, and $s(c)
\equiv \sin(\cos)\theta_W$. 

Equation (1) also generates  new contributions to the 
triple gauge boson vertex. Using the standard parametrization
for the C and P conserving vertex \cite{classical}
\begin{eqnarray}
{\cal L}_{WWV} & = & g_{WWV} \Bigg\{ 
g_1^V \Big( W^+_{\mu\nu} W^{- \, \mu} V^{\nu} 
  - W^+_{\mu} V_{\nu} W^{- \, \mu\nu} \Big) \nonumber\\ 
& + & \kappa_V W_\mu^+ W_\nu^- V^{\mu\nu}
+ \frac{\lambda_V}{M_W^2} W^+_{\mu\nu} W^{- \, \nu\rho} V_\rho^{\; \mu}
 \Bigg\}
\;,
\label{WWV}
\end{eqnarray}
where $V=Z,\gamma$, the coupling constants are 
$g_{WW\gamma} =  e$ and 
$g_{WWZ} = e/(s\,c)$.  The field-strength tensors include only
the Abelian parts, {\em i.e.}\/ 
$W^{\mu\nu} = \partial^\mu W^\nu - \partial^\nu W^\mu$
and $V^{\mu\nu} = \partial^\mu V^\nu - \partial^\nu V^\mu$, and  
\begin{eqnarray}
g_1^Z& = & 1 + \frac{1}{2}\frac{M_Z^2}{\Lambda^2}f_W \;, 
\nonumber \\
\kappa_\gamma & = & 1 + \frac{1}{2}\frac{M_W^2}{\Lambda^2}
\Big(f_W + f_B\Big) \;, \\
\kappa_Z & = & 1 + \frac{1}{2}\frac{M_Z^2}{\Lambda^2}
  \Big(c^2 f_W - c^2 f_B\Big)\;, 
\nonumber \\
\lambda_\gamma = \lambda_Z & = 
& \frac{3}{2} s^2\frac{M_W^2}{\Lambda^2}f_{WWW} \;.
\end{eqnarray}
As seen above, the operators ${\cal O}_{W}$ and ${\cal
O}_{B}$ give rise to both anomalous Higgs--gauge boson couplings
and to new triple and quartic self--couplings amongst the gauge
bosons, while the operator ${\cal O}_{WWW}$ solely modifies the
gauge boson self--interactions.
The operators  ${\cal O}_{WW}$ and ${\cal O}_{BB}$ only affect
$HVV$ couplings, like $HWW$, $HZZ$, $H\gamma\gamma$ and
$HZ\gamma$, since their contribution to the $WW\gamma$ and $WWZ$
tree--point couplings can be completely absorbed in the
redefinition of the SM fields and gauge couplings \cite{hsz}. Therefore, one
cannot obtain any constraint on these couplings from the study of
anomalous trilinear gauge boson couplings. 

\section{New Higgs Signatures}
\label{signals}
In this talk I will review our results on 
Higgs production at the Fermilab
Tevatron collider and at LEPII with its subsequent decay into 
two photons \cite{ourwork}. This channel in the SM occurs at one--loop level
and it is quite small, but due to the new interactions
(1), it can be enhanced and even become dominant. 
I will summarized our results on the signatures:
\begin{eqnarray}
p\, \bar p & \rightarrow & j \, j \, \gamma\, \gamma \nonumber \\
p\, \bar p & \rightarrow & \gamma \,\gamma \,+ \,\not \!\! E_T \nonumber \\
p\, \bar p & \rightarrow & \gamma\, \gamma\, \gamma\\
e^+\, e^-  & \rightarrow & j\, j\, \gamma\, \gamma \nonumber \\
e^+\, e^-  & \rightarrow & \gamma\, \gamma\, \gamma \nonumber
\label{proc}
\end{eqnarray}
 
We have included in our calculations all SM (QCD plus
electroweak), and anomalous contributions that lead to these
final states. The SM one-loop contributions to the $H\gamma\gamma$
and $H Z\gamma$ vertices were introduced through the use of the 
effective operators with the corresponding form factors in the coupling 
\cite{h:rev}. Neither the narrow--width approximation for the Higgs boson 
contributions, nor the effective $W$ boson approximation were employed. We
consistently included the effect of all interferences between the
anomalous signature and the SM background. 
As an example of I quote here that 1928 SM amplitudes plus 236 anomalous ones,
contribute to the process 
$p\, \bar p\rightarrow  j \, j \, \gamma\, \gamma $ \cite{jjgg}. 
The SM Feynman diagrams corresponding to
the background subprocess were generated by Madgraph
\cite{madgraph} in the framework of Helas \cite{helas}. The
anomalous couplings arising from the Lagrangian
(1) were implemented in Fortran routines and were
included accordingly. For the $p\,\bar p$ processes, we have used the 
MRS (G) \cite{mrs} set of
proton structure functions with the scale $Q^2=\hat{s}$.

All processes listed in (7) have been the object of direct
experimental searches. In our analysis we have closely followed
theses searches in order to make our study as realistic as 
possible. In particular when studying the
$\gamma\gamma j j $ final state we have closely followed the
results recently presented by D\O ~Collaboration for $p \bar p
\rightarrow \gamma\gamma j j$ events with high  two--photon
invariant mass \cite{D0:jjgg}. 

For events containing two photons plus large missing transverse
energy ($\gamma\gamma \not \!\! E_T$) as well as 
three photons in the final state we have used the results from 
D\O ~and CDF collaborations \cite{prl:old,prl:new,cdf}.
These events represent an important
signature for some classes of supersymmetric models and in
Refs.\cite{prl:old,prl:new,cdf} the experimental collaborations
use their results to set limits in some of the SUSY parameters. 
However, as we pointed out \cite{ourwork}, these final states can also 
be a signal of Higgs production and subsequent decay into photons and
can be used to place limits on the coefficients of the anomalous operators
(1).
 
Finally, in order to obtain constraints on the anomalous couplings
described above, we have also used the OPAL
data \cite{opal:gg,opal:ggg} for the reactions,
\begin{eqnarray}
e^+ e^- &\rightarrow& \gamma \gamma \gamma \; ,
\label{ggg}
\\
e^+ e^- &\rightarrow& \gamma \gamma + \mbox{ hadrons} \; .
\label{ggh}
\end{eqnarray}

As an example, I describe below more in detail our analysis of the
$\gamma\gamma j j $ final state. 

\section{The process $p\, \bar p\rightarrow j \, j \, \gamma\, \gamma$: An
Example}

As mentioned before when studying the
$\gamma\gamma j j $ final state we have closely followed the
results presented by D\O ~Collaboration for $p \bar p
\rightarrow \gamma\gamma j j$ events with high  two--photon
invariant mass \cite{D0:jjgg}. 
The cuts applied on the final state particles are: 
\begin{displaymath}
\begin{array}{ll}
\mbox{For the photons} & \\[0.1cm]
|\eta_{\gamma 1}|< 1.1 \mbox{  or  } 1.5<|\eta_{\gamma 1}|<2 & 
p_T^{\gamma 1}>20 \mbox{ GeV} \\
|\eta_{\gamma 2}|< 1.1 \mbox{  or  } 1.5<|\eta_{\gamma 2}|<2.25 & 
p_T^{\gamma 2}>25 \mbox{ GeV} \\
\sum \vec p_T^\gamma >10 \mbox{ GeV} &  \\[0.1cm]
\mbox{For the $l \nu \gamma\gamma$ final state} &\\[0.1cm] 
|\eta_{e}|< 1.1 \mbox{  or  } 1.5<|\eta_e|<2 & |\eta_{\mu}|< 1 \\
p_T^{e,\mu}>20 \mbox{ GeV} &{\not \!p}_T > 20 \mbox{ GeV} \\ [0.1cm]
\mbox{For the $j j \gamma\gamma$ final state} & \\ [0.1cm]
|\eta_{j 1}|< 2  & 
p_T^{j 1} > 20 \mbox{ GeV} \\
|\eta_{j 2}|< 2.25 & 
p_T^{j2}> 15 \mbox{ GeV} \\
\sum \vec p_T^j >10 \mbox{ GeV} & R_{\gamma j} > 0.7  \\
40\le M_{jj} \le 150 \mbox{ GeV}
\end{array}
\end{displaymath} 
We also assumed an invariant--mass resolution for the
two--photons of  $\Delta M_{\gamma\gamma}/ M_{\gamma\gamma} =
0.15/ \sqrt{M_{\gamma\gamma}} \oplus 0.007$ \cite{h:smw}. Both
signal and background were integrated over an invariant--mass bin
of $\pm 2 \Delta M_{\gamma\gamma}$ centered around $M_H$.
Finally,we isolate the majority
of events due to associated production, and the corresponding
background, by integrating over a bin centered on the $W$ or $Z$
mass, which is equivalent to the invariant mass cut listed above. 

After imposing all the cuts, we get a reduction on the signal
event rate which depends on the Higgs mass. For instance, for the
$jj \gamma\gamma$ final state the geometrical acceptance and
background rejection cuts account for a reduction factor of 15\% for
$M_H=60$ GeV rising to 25\% for $M_H=160$ GeV. We also include in
our analysis the particle identification and trigger
efficiencies. For leptons and photons they vary from 40\% to 70\%
per particle \cite{D0,D02}. For the $jj\gamma\gamma$ final
state we estimate the total effect of these  efficiencies to be
35\%. We therefore obtain an overall efficiency for the
$jj\gamma\gamma$ final state of 5.5\% to 9\% for $M_H =
60$--$160$ GeV in agreement with the results of Ref.\
\cite{D0:jjgg}. 

Dominant backgrounds are due to missidentification
when a jet fakes a photon. The probability for a jet to fake a
photon has been estimated to be of a few times $10^{-4}$
\cite{D0}. Although this probability is small, it becomes the
main source of  background for the $j j \gamma\gamma$ final state
because of the very large multijet cross section. In Ref.\
\cite{D0:jjgg} this background is estimated to lead to $3.5\pm
1.3$ events with invariant mass  $M_{\gamma\gamma}>60 $ GeV and
it has been consistently included in our derivation of the
attainable limits. 

\section{Results and Conclusion}
\label{res:con}

I now present our results on the attainable limits on the 
coefficients of the anomalous operators. In order to establish these
bounds on the coefficients in each process,
we imposed an upper limit on the number of signal events 
based on Poisson statistics. In the absence of background 
this implies
$N_{\mbox{signal}} < 1 \,(3)$ at 64\% (95\%) CL. In the presence
of background events, we employed the modified Poisson analysis. 
We are currently working on the statistical
combination of the information from the different final states
\cite{preparation}. 

\begin{figure}
\begin{center}
\mbox{\psfig{file=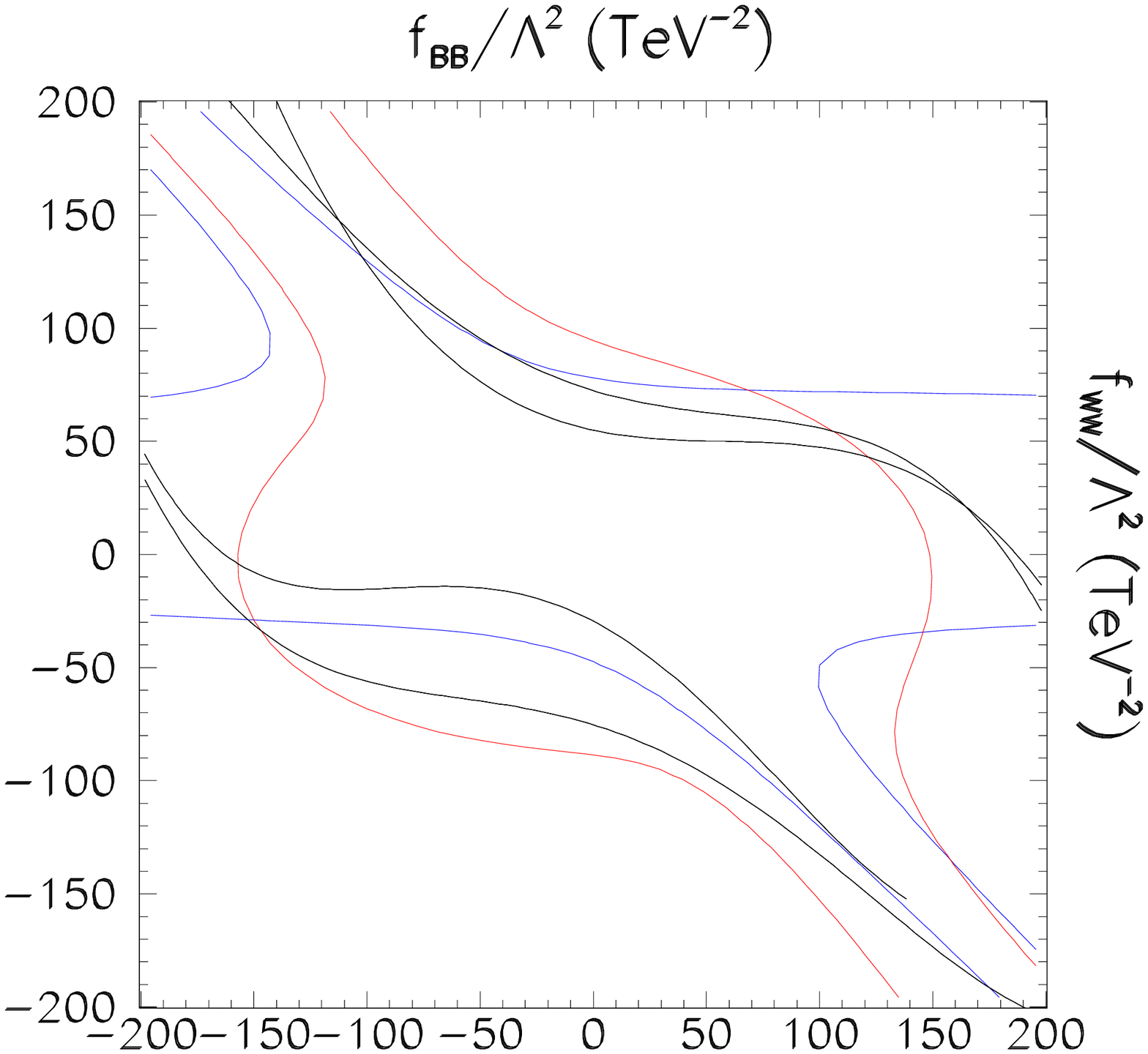,width=0.3\textwidth}}
\mbox{\psfig{file=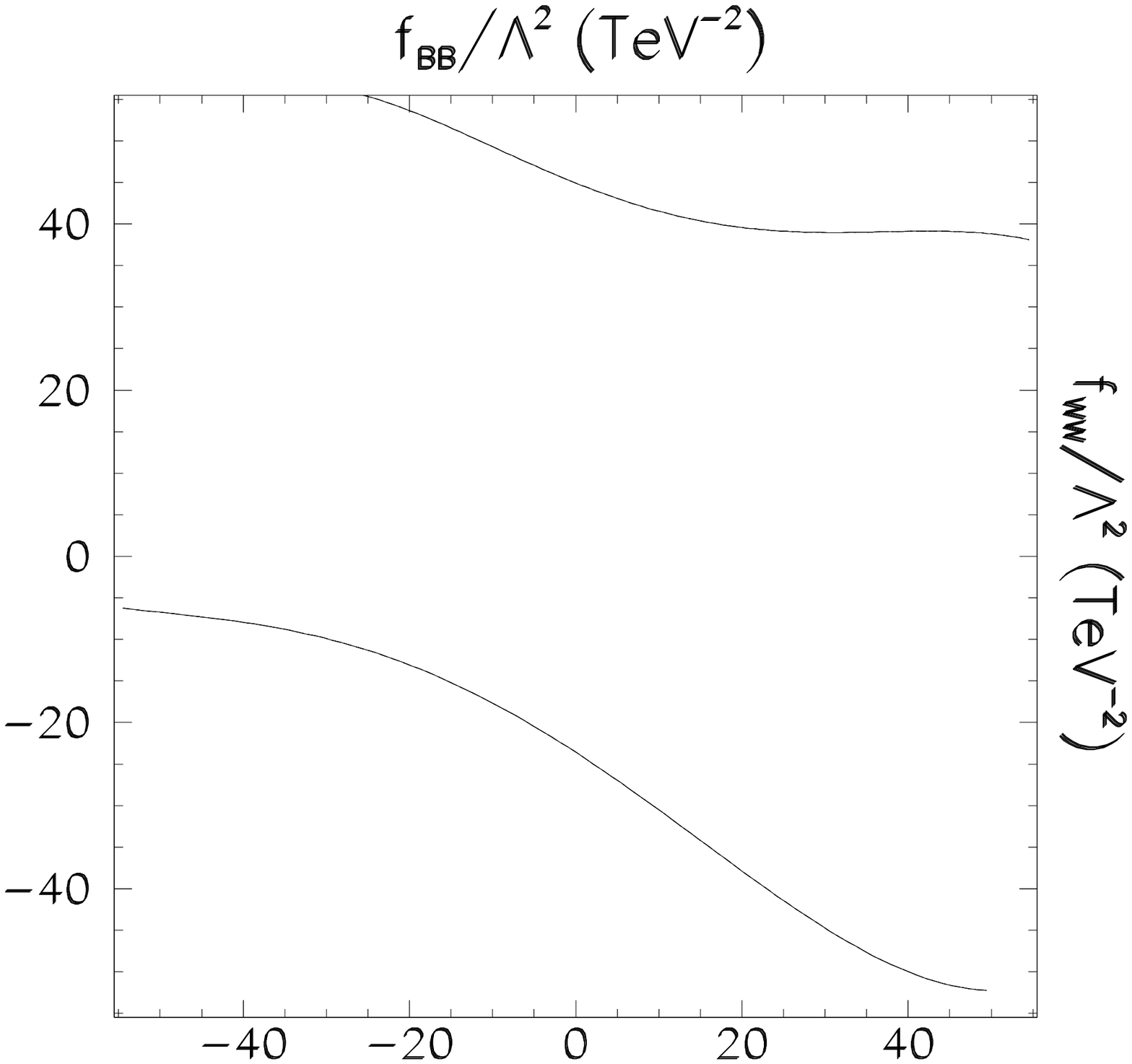,width=0.3\textwidth}}
\end{center}
\caption{{\bf a)}Exclusion region outside the curves in the  $f_{BB}
\times f_{WW}$ plane, in TeV$^{-2}$, based on the CDF analysis
of $\gamma\gamma\gamma$ production (most external 
black lines), on the D0 analysis of $\gamma\gamma j j$ 
production (most internal black lines), on the D0 analysis 
of $\gamma\gamma \not \!\! E_T$ (blue lines), and on the OPAL  analysis 
of $\gamma\gamma\gamma$ production (red lines)
, always assuming $ M_H = 100$ GeV.  The curves show the 95\%
CL deviations from the SM total cross section.
{\bf b)} Same as {\bf a)} for all processes combined.}
\label{contour}
\end{figure}

The coupling $H\gamma\gamma$ derived from (2)
involves $f_{WW}$ and $f_{BB}$ \cite{hsz}.  In consequence, the
anomalous signature $f\bar f \gamma\gamma$ is only possible when
those couplings are not vanishing. The couplings $f_B$ and $f_W$,
on the other hand, affect the production mechanisms for the Higgs
boson. 
In Fig.\ \ref{contour}.a we present our results for the excluded 
region in the $f_{WW}$, $f_{BB}$ plane from the different channels 
studied \cite{ourwork} for $M_H=100$ GeV assuming that these
are the only non-vanishing couplings. Since the anomalous contribution to 
$H\gamma\gamma$ is zero for
$f_{BB} = - f_{WW}$, the bounds become very weak close to this
line, as is clearly shown in Fig.\ \ref{contour}. 
In Fig.\ \ref{contour}.b we show the preliminary results 
for the same plot after combining all the channels. 
As seen in the figure, one expects a clear
improvements of the individual bounds, when the information from all 
channels is combined.

These bounds depend on the Higgs mass and became weaker
as the Higgs boson becomes heavier. In Table \ref{tab:f1} we 
display the allowed values for $f/\Lambda^2$, 
at 95\% CL, from  $\gamma\gamma j j $
Tevatron D0 data analysis assuming that $f_{WW}=f_{BB}$  
and $f_W=f_B=0$ for different Higgs masses. For the sake of completeness
we also show the accessible bound for future Tevatron Upgrades.
We should remind that this scenario will not be
restricted by data on $W^+W^-$ production since there is
no trilinear vector boson couplings involved. Therefore the limits
here presented are the only existing direct bounds on these operators. 
\begin{table*}
\caption{Allowed range of $f/\Lambda^2$ in TeV$^{-2}$ at 95\% CL,
assuming that ($f_{BB}=f_{WW}\gg f_B, f_W$) for the
different final states, and for different Higgs boson masses}
\begin{tabular}{|c|c|c|c|c|c|}
\hline
 $M_H$ (GeV) &        & 100 & 150 & 200 & 250  \\
\hline
$j \, j \, \gamma \, \gamma$ &   RunI &
                             ($-20$ --- 49) & 
                             ($-26$ --- 64) &
                             ($-96$ ---$>100 $) &  
                             ($<-100$ --- $>100$)   
                             \\
&  RunII &
                             ($-8.4$ --- 26) & 
                             ($-11$ --- 31) &
                             ($-36$ --- 81) &  
                             ($-64$ --- $>100$) \\ 
&  TeV33 &
                             ($-4.2$ --- 6.5) & 
                             ($-4.5$ --- 12) &
                             ($-19$ --- 40) &  
                             ($-28$ --- 51) \\ \hline
\end{tabular}
\label{tab:f1}
\end{table*}

\begin{figure}
\begin{center}
\mbox{\psfig{file=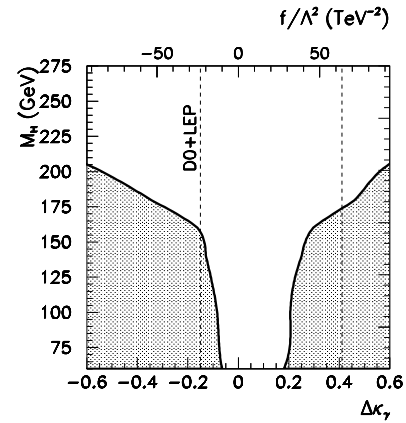,width=0.3\textwidth}}
\end{center} 
\caption{Excluded region in the $\Delta\kappa_\gamma \times M_H$
plane from the combined analysis 
from  the combined results of the  $\gamma\gamma\gamma$ 
production at LEP, $\gamma\gamma\gamma$, $\gamma\gamma +  \!\! E_T $, and
$\gamma\gamma j j $ production at Tevatron, assuming that all $f_i$ are equal
(see text for details).}
\label{fig:2}
\end{figure}

One may wonder how reasonable are these bounds, or how they compare
with other existing limits on the coefficients of other dimension-six
operator. In order to address this question one can
make the assumption that all blind operators affecting the Higgs
interactions have a common coupling $f$~\cite{hisz}, {\it i.e.}
\begin{equation}
f_W = f_B = f_{WW} = f_{BB} = f \; , 
\label{blind}
\end{equation}
In this scenario, $g^{(1)}_{H Z
\gamma} = g^{(3)}_{HZZ} = 0$, and we can relate the Higgs boson
anomalous coupling $f$ with the conventional parametrization of
the vertex $WWV$ ($V=Z$, $\gamma$) 
\begin{equation}
\Delta \kappa_\gamma = \frac{M_W^2}{\Lambda^2}~ f=c^2 
\Delta g^Z_1\;
\;\;,
\Delta \kappa_Z = \frac{( 1 - 2 s^2)}{c^2}\Delta\kappa_\gamma
\;\; .
\label{trad} 
\end{equation}
In Fig.\ \ref{fig:2}, we show the region in
the $\Delta \kappa_\gamma \times M_H$ that can be excluded
through the combined analysis of the  $\gamma\gamma\gamma$ 
production at LEP, $\gamma\gamma\gamma$, $\gamma\gamma +  \!\! E_T $, and
$\gamma\gamma j j $ production at Tevatron \cite{preparation}.
For the sake of comparison, we also show in Fig.\ \ref{fig:2} the
best available  experimental limit on $\Delta \kappa_\gamma$
from double gauge
boson production at Tevatron and LEP II \cite{vancouver}.
In all cases the results were obtained
assuming the HISZ scenario. We can see that, for $M_H \leq 
170$ GeV, the limit that can be established at 95\% CL from
the Higgs production analysis is tighter than the
present limit coming from gauge boson production. 

In conclusion, we have shown that the  analysis
of an anomalous Higgs boson production at the Fermilab
Tevatron and the CERN LEP II collider may be used to impose
strong  limits on new effective interactions. Under the
assumption that the coefficients of the four ``blind'' effective
operators contributing to Higgs--vector boson couplings are of the
same magnitude, the study can give rise to a significant indirect
limit on anomalous $WW\gamma$ couplings. Furthermore, this
analysis  is able to set constraints on those operators
contributing to new Higgs interactions for Higgs masses far
beyond the kinematical reach of LEP II. 


\end{document}